\documentclass{PoS}

\usepackage{amsmath , amsthm , amssymb }
\usepackage{braket}
\usepackage{dsfont}
\usepackage{feynmp}
\usepackage{graphicx}
\usepackage{tikz}

\title{Calculation of the Nucleon Axial Form Factor Using Staggered Lattice QCD}

\ShortTitle{Axial Form Factor with HISQ}

\author{\speaker{Aaron S. Meyer}\\
        Enrico Fermi Institute and Department of Physics,\\
          The University of Chicago, Chicago, Illinois, 60637, USA and\\
          Fermi National Accelerator Laboratory, Batavia, Illinois, USA\\
        E-mail: \email{asmeyer2012@uchicago.edu}}

\author{Richard J. Hill\\
        TRIUMF, 4004 Wesbrook Mall, Vancouver, British Columbia, V6T 2A3 Canada\\
          Perimeter Institute for Theoretical Physics, Waterloo, Ontario, N2L 2Y5 Canada and\\
          Enrico Fermi Institute and Department of Physics,\\
          The University of Chicago, Chicago, Illinois, 60637, USA\\
        E-mail: \email{richardhill@uchicago.edu}}

\author{Andreas S. Kronfeld\\
        Fermi National Accelerator Laboratory,
          Batavia, Illinois, USA and\\
          Institute for Advanced Study, Technische Universit\"at M\"unchen,
          85748 Garching, Germany\\
        E-mail: \email{ask@fnal.gov}}

\author{Ruizi Li\\
        Department of Physics, Indiana University,
          Bloomington, Indiana, USA\\
        E-mail: \email{ruizli@umail.iu.edu}}

\author{James N. Simone\\
        Fermi National Accelerator Laboratory,
          Batavia, Illinois, USA\\
        E-mail: \email{simone@fnal.gov}}

\abstract{
The nucleon axial form factor is a dominant contribution to errors in neutrino oscillation studies.
Lattice QCD calculations can help control theory errors
 by providing first-principles information on nucleon form factors.
In these proceedings, we present preliminary results on a blinded calculation of $g_A$
 and the axial form factor using HISQ staggered baryons with 2+1+1 flavors of sea quarks.
Calculations are done using physical light quark masses and are absolutely normalized.
We discuss fitting form factor data with the model-independent $z$~expansion parametrization.
}

\FullConference{34th annual International Symposium on Lattice Field Theory\\
		24-30 July 2016\\
		University of Southampton, UK}

\begin{document}
\DeclareGraphicsRule{*}{mps}{*}{}

\section{Introduction}
\label{sec:intro}

Neutrino physics is in a golden age of discovery.
Physicists aim to determine many elusive neutrino properties, such as the neutrino mass ordering,
 the possible existence of sterile neutrinos, the value of $\delta_{CP}$,
 and to make a precision measurement of $\theta_{23}$.
Future neutrino oscillation experiments could even test for unitarity violation in the
 lepton sector.
Next generation neutrino oscillation experiments are poised to probe these exciting new questions,
 but contributions from theory are vital to the success of experiment goals.

To measure neutrino oscillation parameters precisely,
 one must have sufficiently precise knowledge of the neutrino cross sections
 for the nuclear targets used in the experiment.
While targets with fewer nucleons are ideal from a systematics standpoint,
 these targets are often impractical because of the tiny cross sections involved.
As a consequence, large nuclear targets are employed and experiments rely on
 accurate predictions of nuclear properties from modeling.

While nuclear models are often blamed as the primary contribution to theory systematics
 in oscillation experiments, free nucleon amplitudes, which are input into these
 nuclear models, are also a cause for concern.
Typical parametrizations for the form factors are poorly justified and underestimate errors.
Furthermore, the axial form factor is most directly probed by neutrino scattering
 for which there is a sparsity of data on elementary targets.
Since this form factor is part of the leading contribution to nucleon amplitudes,
 a robust determination of the form factor is a priority.
Our goal is to improve the free nucleon amplitudes using lattice QCD.

In these proceedings, we discuss the determination of the nucleon axial form factor.
In Sec.~\ref{sec:neutrino} we discuss recently reanalyzed data from past neutrino
 bubble chamber experiments using a model-independent parametrization of the axial form factor.
This analysis provides realistic errors corresponding to our current best knowledge of
 the axial form factor.
In Secs.~\ref{sec:lattice} and~\ref{sec:2pt} we describe a first-principles
 computation of the axial form factor using the HISQ action.
A first look at data for a HISQ calculation of $g_A$ is presented in Sec.~\ref{sec:3pt}.

\section{Nucleon Form Factors}
\label{sec:neutrino}

There are four nucleon form factors which are relevant for free-nucleon scattering;
 the Dirac and Pauli form factors $F_{1}$ and $F_{2}$,
 the axial form factor $F_A$,
 and the pseudoscalar form factor $F_P$~\cite{Formaggio:2013kya}.
The vector form factors are determined from high-statistics experiments using
 electron scattering off of proton targets.
The pseudoscalar form factor is related to the axial form factor via the 
 Partially Conserved Axial Current (PCAC) condition and its effects are suppressed
 by the lepton mass~\cite{Adler:1964yx}.
The axial form factor affects neutrino cross sections at the same level as the
 vector form factors, but is only determined from low-statistics neutrino scattering experiments.
The nucleon axial form factor is thus the largest contributor to the
 systematic errors and is the focus of this study.
The other form factors can and will be calculated with lattice QCD as consistency checks.

The neutrino community typically assumes the axial form factor $Q^2$ dependence
 has a dipole shape~\cite{LlewellynSmith:1971uhs}
\begin{equation}
F_A(Q^2) = \frac{g_A}{(1+Q^2/M_A^2)^2}
\end{equation}
 where $M_A$ is a free parameter and $g_A$ is taken from neutron $\beta$ decay.
We advocate the model-independent $z$~expansion~\cite{Bhattacharya:2011ah},
where $z$ is related to $t=-Q^2$ by a conformal mapping.
The $z$~expansion mapping is contained in the equation
\begin{equation}
z(t;t_0,t_c) = \frac{\sqrt{t_c-t} - \sqrt{t_c - t_0}}{\sqrt{t_c-t} + \sqrt{t_c-t_0}}\,,
\end{equation}
with $t_c=9m_\pi^2$ and $t_0$ chosen to optimize convergence of the expansion over some
 interesting kinematic region.
The $z$~expansion is simply a power series in $z$,
\begin{equation}
F_A(Q^2(z)) = \sum_{k=0}^{\infty} a_k z^k\,,
\end{equation}
 which converges for all $|z|<1$ owing to unitarity constraints.

We recently reanalyzed deuterium bubble chamber data, comparing the dipole axial form
 factor with the $z$~expansion~\cite{Meyer:2016oeg}.
Three data sets were used, all of $O(1000)$ events.
The results show that using the dipole underestimates the error on the cross section
 by as much as an order of magnitude.
For example, Ref.~\cite{Meyer:2016oeg} finds the cross section 
 at $E_\nu=1~{\rm GeV}$ using the $z$~expansion is $10.1(0.9)\times10^{-39}~{\rm cm}^2$
 compared with $10.63(0.14)\times10^{-39}~{\rm cm}^2$
 using the dipole form factor with $M_A=1.014(14)~{\rm GeV}$ from Ref.~\cite{Bodek:2007ym}.

\section{Formalism for HISQ Spectrum Calculation}
\label{sec:lattice}
Another way to obtain the axial form factor is to calculate it with lattice QCD.
We are calculating $F_A(Q^2)$ using staggered quarks
 on the MILC HISQ 2+1+1 gauge ensembles~\cite{Bazavov:2012xda}.
This choice of action confers several advantages.
There is no explicit chiral symmetry breaking in the $m\rightarrow 0$ limit, and thus
 no exceptional configurations.
Staggered quarks are computationally fast, enabling high statistics, large volumes,
 and physical pion mass at several lattice spacings.
As discussed below, it is straightforward to absolutely normalize the axial current.
The chiral symmetry and absolute normalization simplify interpretation of the results.
The large volumes reduce contributions from finite size effects,
 and with physical-mass pions we can use chiral perturbation theory to correct for
 those that remain.

A few disadvantages to using staggered quarks must be addressed.
These problems include the complicated group theory of staggered fermions and
 the presence of many baryon tastes
 (extra states which are the result of lattice artifacts, analogous to flavors)
 in correlation functions.
The staggered quark group was studied for baryons in detail by
 Golterman and Smit~\cite{Golterman:1984dn},
 Kilcup and Sharpe~\cite{Kilcup:1986dg}, and Bailey~\cite{Bailey:2006zn}.
Building on the decomposition of Kilcup and Sharpe,
 the staggered symmetries of a time slice can be written as the group
\begin{equation}
(((\mathcal{T}_M\times \mathds{Q}_8) \rtimes W_3)\times D_4)/\mathds{Z}_2\,.
\end{equation}
Here $\rtimes$ denotes a semidirect product.
This decomposition separates the lattice translations 
 $\mathcal{T}_M \cong (\mathds{Z}_{N})^3$ and
 the rotations $W_3$ (the octahedral group)
 from the discrete taste transformations $\mathds{Q}_8$ and $D_4$
 (the order-8 quaternion group and the order-8 dihedral group, respectively).
The quotient factor $\mathds{Z}_2$ identifies the double cover in the $D_4$ taste
 group and the $W_3$ rotation group as the same.

This group has three fermionic irreps: $8$, $8^\prime$, and $16$,
 named according to the dimension of the irrep.
The staggered field transforms under the $8$ irrep, with operators of higher spin 
 showing up in the $8^\prime$ and $16$ irreps.
The operators transforming under the lattice symmetries generate ``$N$-like'' and ``$\Delta$-like''
 states.
The mass of an ``$N$-like'' state converges to the nucleon mass in the continuum limit,
 regardless of whether the operators transform with isospin $\frac{1}{2}$ or $\frac{3}{2}$.
Similarly, the mass of a ``$\Delta$-like'' state converges to the $\Delta$ mass in the
 continuum limit.

The way these states appear in each irrep and isospin combination was
 studied in detail by Bailey~\cite{Bailey:2006zn}.
He found that the number of operators available for each irrep is equal to the number of
 lowest-order baryon taste states that couple to that irrep.
This means that we can always construct an operator basis with enough information
 to extract all of the lowest-order states via a variational method.
In practice, we use a fit to multiple exponentials with Bayesian priors and are able
 to extract more than just the lowest-order states guaranteed by the variational method.

There are a variety of permissible baryon operators for staggered quarks,
 as outlined by Golterman and Smit~\cite{Golterman:1984dn} and Bailey~\cite{Bailey:2006zn}.
We compute all nonvanishing combinations of operators with isospin $\frac{3}{2}$ to
 investigate the nucleon properties.

\section{Two-Point Correlation Functions}
\label{sec:2pt}

\begin{figure}[t]
\centering
$
\begin{array}{ccc}
\begin{tikzpicture}
\node at (0,0) {\includegraphics[width=0.45\textwidth]{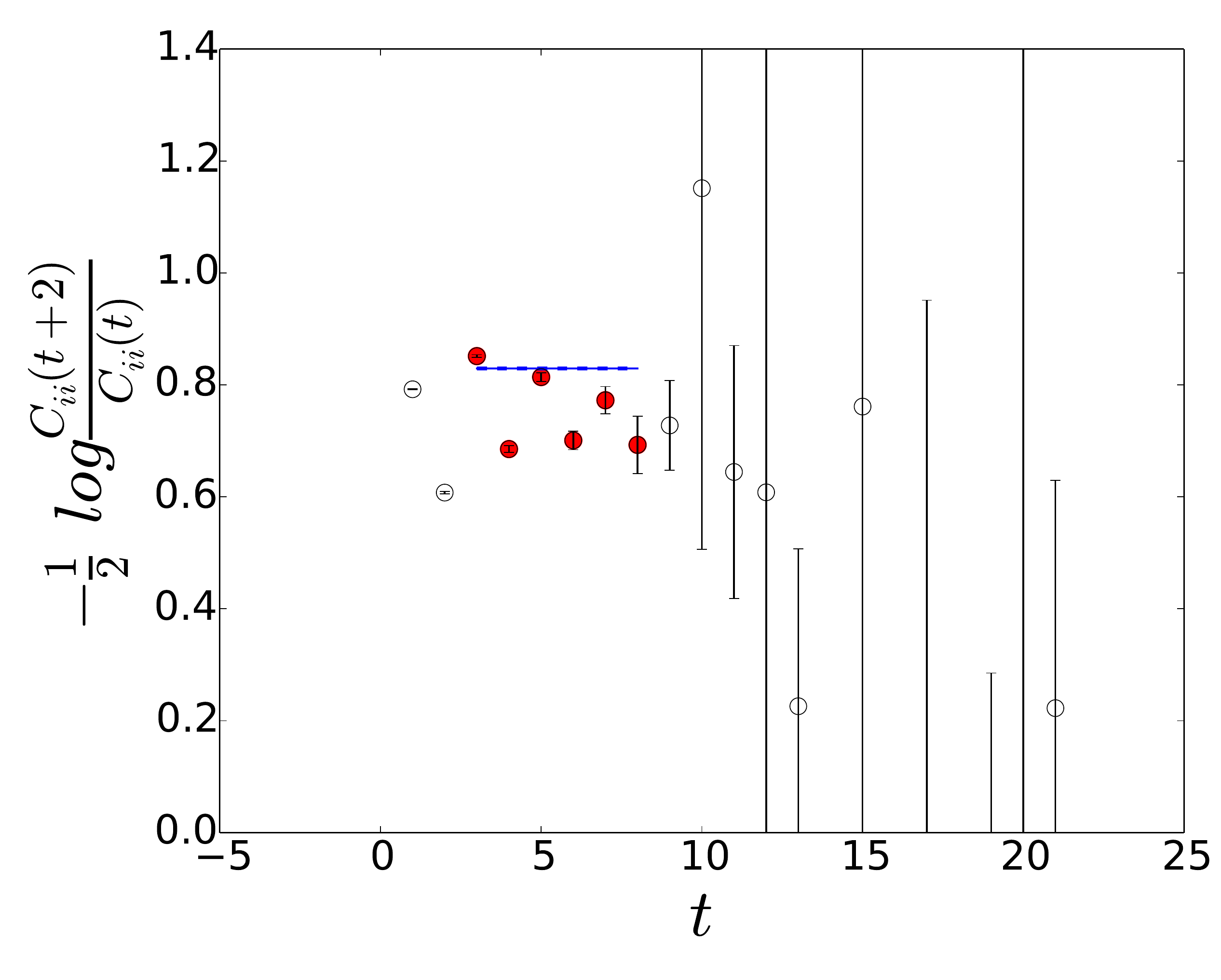}};
\node at (-1.1,2.2) {\textcolor{red}{\footnotesize PRELIMINARY}};
\end{tikzpicture} &&
\begin{tikzpicture}
\node at (0,0) {\includegraphics[width=0.45\textwidth]{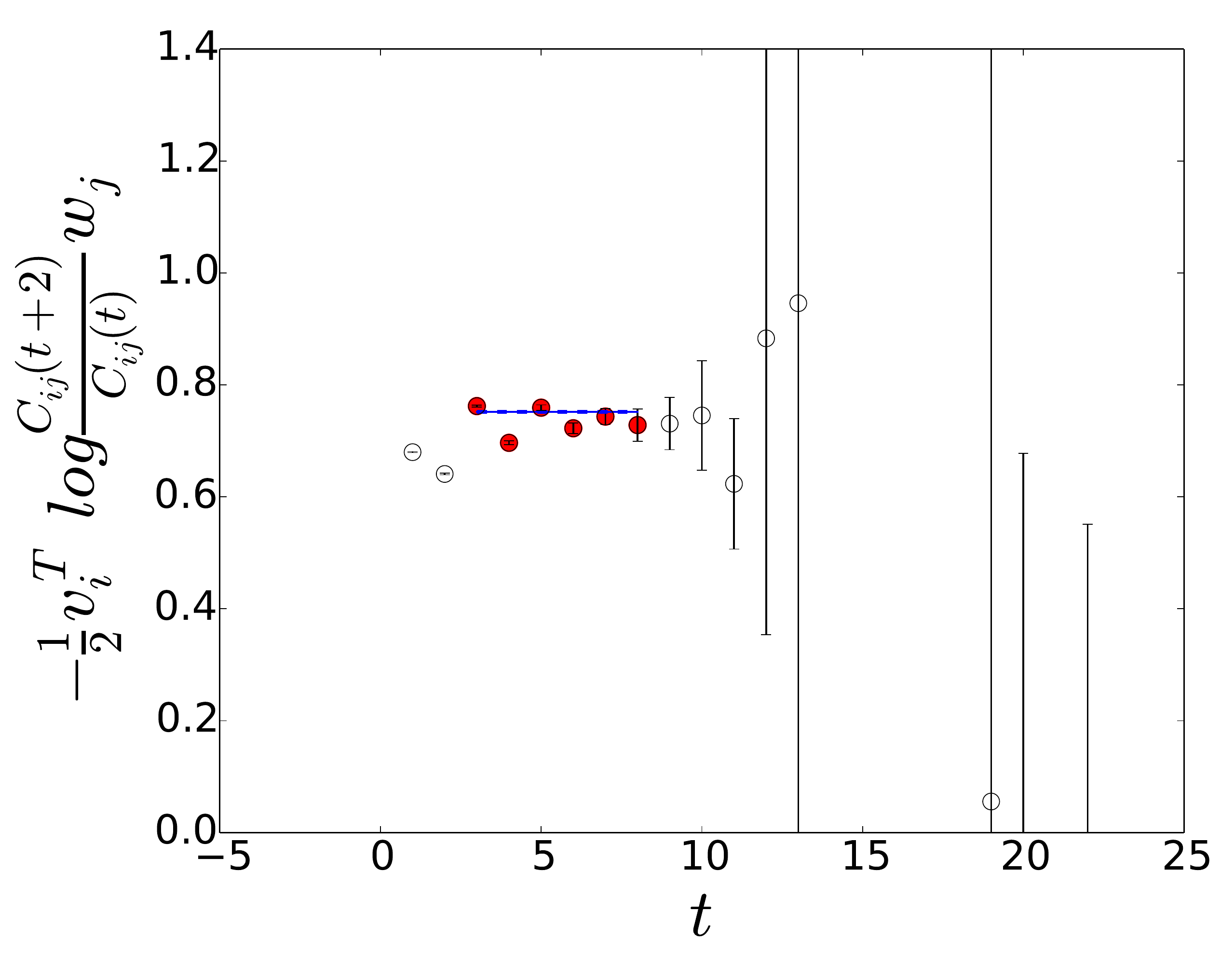}};
\node at (-1.1,2.2) {\textcolor{red}{\footnotesize PRELIMINARY}};
\end{tikzpicture}
\end{array}
$
\caption{
 \label{fig:2pteffmass}
 Effective mass plots for the two-point correlation function data.
 The solid red points are those which have been used to fit a plateau region.
 Left: A typical unoptimized effective mass plot.
 Right: The signal to noise optimization in
  Eq.~\protect\ref{eq:2ptsnopt} has been applied.
 The effective mass more strongly resembles an effective mass from just the less-noisy $N$-like
  state.
}
\end{figure}

We first look at the effective mass of the raw correlation functions, as shown in 
 Fig.~\ref{fig:2pteffmass}.
A plateau is hard to see.
We therefore optimize a metric related to the signal to noise by varying $v$ and $w$
 in the equation
\begin{equation}
\frac{S^2}{N^2} = \sum_{ij}\sum_{t=t_{\text{min}}}^{t_{\text{max}}}
 \frac{\left[v_i^T \mathcal{E}_{ij}(t) w_j\right]^2}
 {\,\,\delta\left[v_i^T \mathcal{E}_{ij}(t) w_j\right]^2}\,.
 \label{eq:2ptsnopt}
\end{equation}
 to maximize $S^2/N^2$.
Here, $\mathcal{E}_{ij}(t)$ denotes the effective mass for operator
 source $i$ and sink $j$ at time $t$ and $\delta\left[\dots\right]^2$ denotes the square
 of the error on the quantity in the brackets.
This metric is a tool which allows us to better understand correlations,
 although final fits do not depend on this optimization.
The result of optimizing this metric for the effective mass is shown in
 Fig.~\ref{fig:2pteffmass}.

\begin{figure}[t]
$
\begin{array}{ccc}
\begin{tikzpicture}
\node at (0,0) {\includegraphics[width=0.45\textwidth]{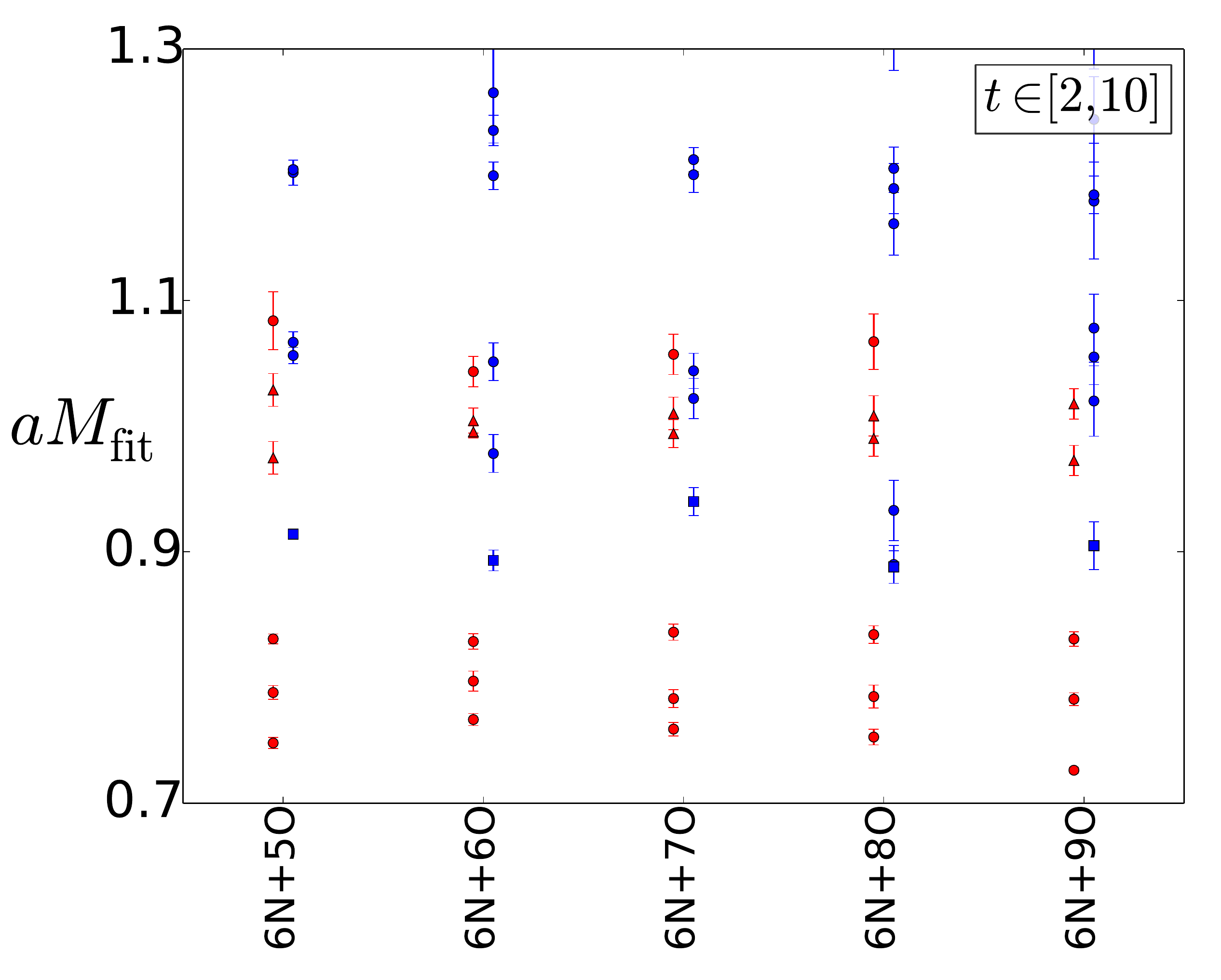}};
\node at (-1.2,2.2) {\textcolor{red}{\footnotesize PRELIMINARY}};
\end{tikzpicture}
 &&
\begin{tikzpicture}
\node at (0,0) {\includegraphics[width=0.45\textwidth]{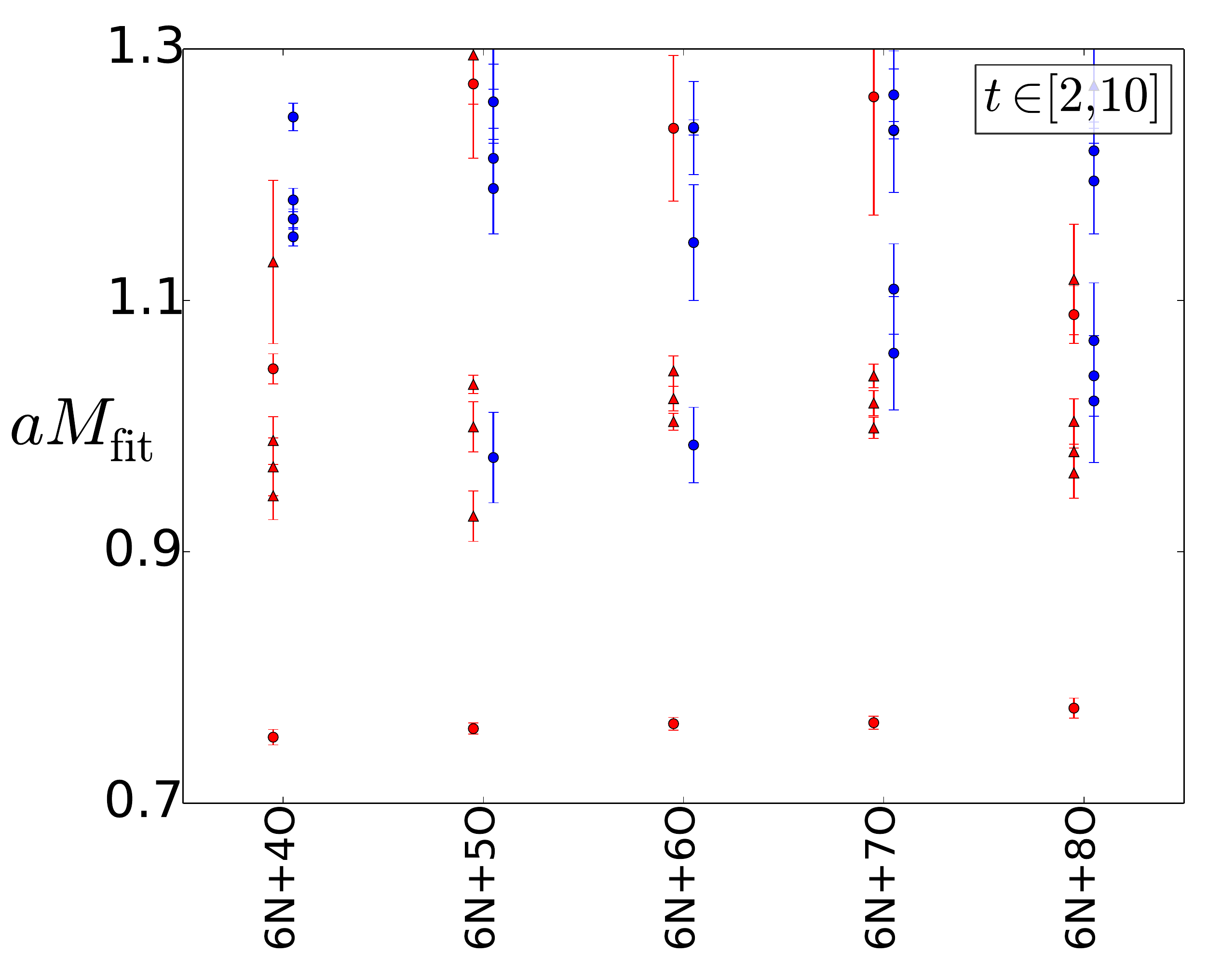}};
\node at (-1.2,2.2) {\textcolor{red}{\footnotesize PRELIMINARY}};
\end{tikzpicture}\\
\end{array}
$
\caption{
 \label{fig:2ptstability}
 Stability plots for the $8$ (left) and $16$ (right) irreps.
 The spectra shown are for an ensemble with $a\approx 0.15~{\rm fm}$,
  $L^3\times T = 32^3\times 48$, $M_\pi L\approx 3.4$, and about 1500 measurements.
 The fits shown include a fixed number of even states and increasing number 
  of odd states when moving left to right.
 The non-oscillating states are shown in red (slight left offset) and the oscillating
  parity-partner states are shown in blue (slight right offset).
 $N$-like states are represented with circles and $\Delta$-like states with triangles.
 The error bars shown are the errors on the masses themselves.
 These are highly correlated between states, and errors on the mass splittings
  are significantly smaller.
}
\end{figure}

We now turn to fitting the two-point function data.
Fits are performed to a tower of exponentials with Bayesian priors on the fit parameters.
We fit first to the $8^\prime$ irrep, then use the fit posteriors to inform
 the priors on taste and $N$-$\Delta$ mass splittings for the fit to the $8$ irrep.
The final $16$ irrep fit uses posteriors from both $8$ and $8^\prime$ as mass splitting priors,
 which has only one $N$-like taste. 
Wide priors are used for the $N$-like masses to prevent biasing the results.
Stability plots are shown in Fig.~\ref{fig:2ptstability}.
The fit mass spectrum for the $N$-like states are stable as the number of oscillating
 states is increased.
The $N$-like mass from the $6+7$ state $16$ fit is 999(7)~{\rm MeV} with statistical error only.

\section{Three-Point Correlation Functions}
\label{sec:3pt}

To obtain the axial charge, we plan to compute a ratio of matrix elements
\begin{equation}
\label{eq:3ptratio0mom}
\frac{\bra{N}Z_A A_\mu\ket{N}}{\bra{0}Z_A A_\mu\ket{\pi^a}}\Biggr|_{q=0} \propto
 \frac{g_A}{f_\pi} \,,
\end{equation}
such that the (re)normalization factor cancels out.
We then plan to use $f_\pi$ determined from MILC's computation with the Goldstone pseudoscalar
 density~\cite{Bazavov:2014wgs}.
We blind the computation by multiplying the three-point matrix element by a constant prefactor.
At nonzero momentum, the axial form factor can be computed as a ratio of three-point functions
\begin{equation}
\frac{\bra{N(-Q)}Z_A A_{\perp\mu}(Q)\ket{N(0)}}{\bra{N(0)}Z_A A_\mu(0)\ket{N(0)}} \propto
 \frac{F_A(Q^2)}{g_A}\,,
\end{equation}
where $A_{\perp\mu}$ is the component orthogonal to $Q$:
\begin{equation}
A_{\perp\mu}(Q) = A_{\mu}(Q) - Q_{\mu} \frac{Q\cdot A(Q)}{Q^2}\,.
\end{equation}

\begin{figure}[t]
$
\begin{array}{ccc}
\begin{tikzpicture}
\node at (0,0) {\includegraphics[width=0.45\textwidth]{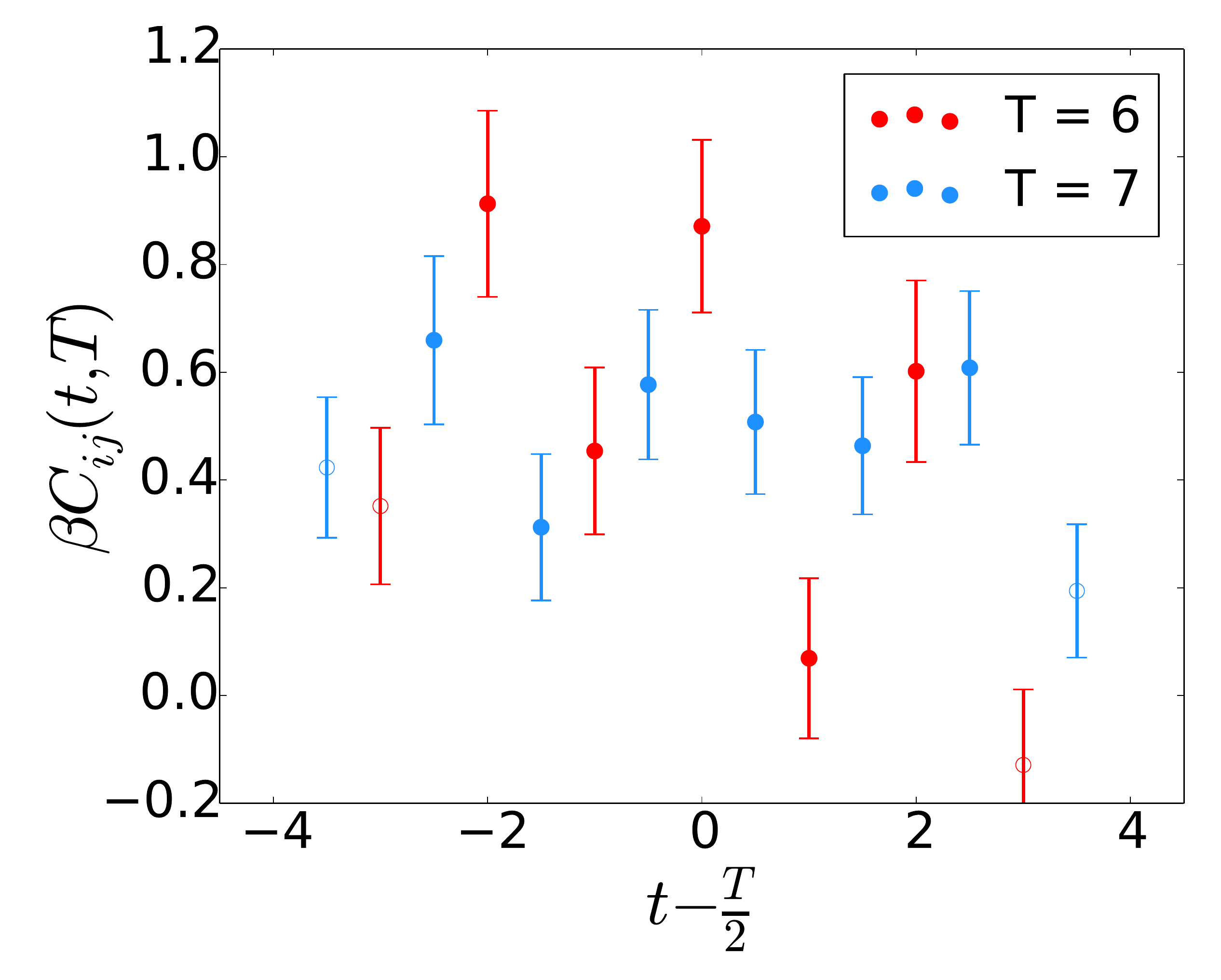}};
\node at (-1.1,2.2) {\textcolor{red}{\footnotesize PRELIMINARY}};
\end{tikzpicture} &&
\begin{tikzpicture}
\node at (0,0) {\includegraphics[width=0.45\textwidth]{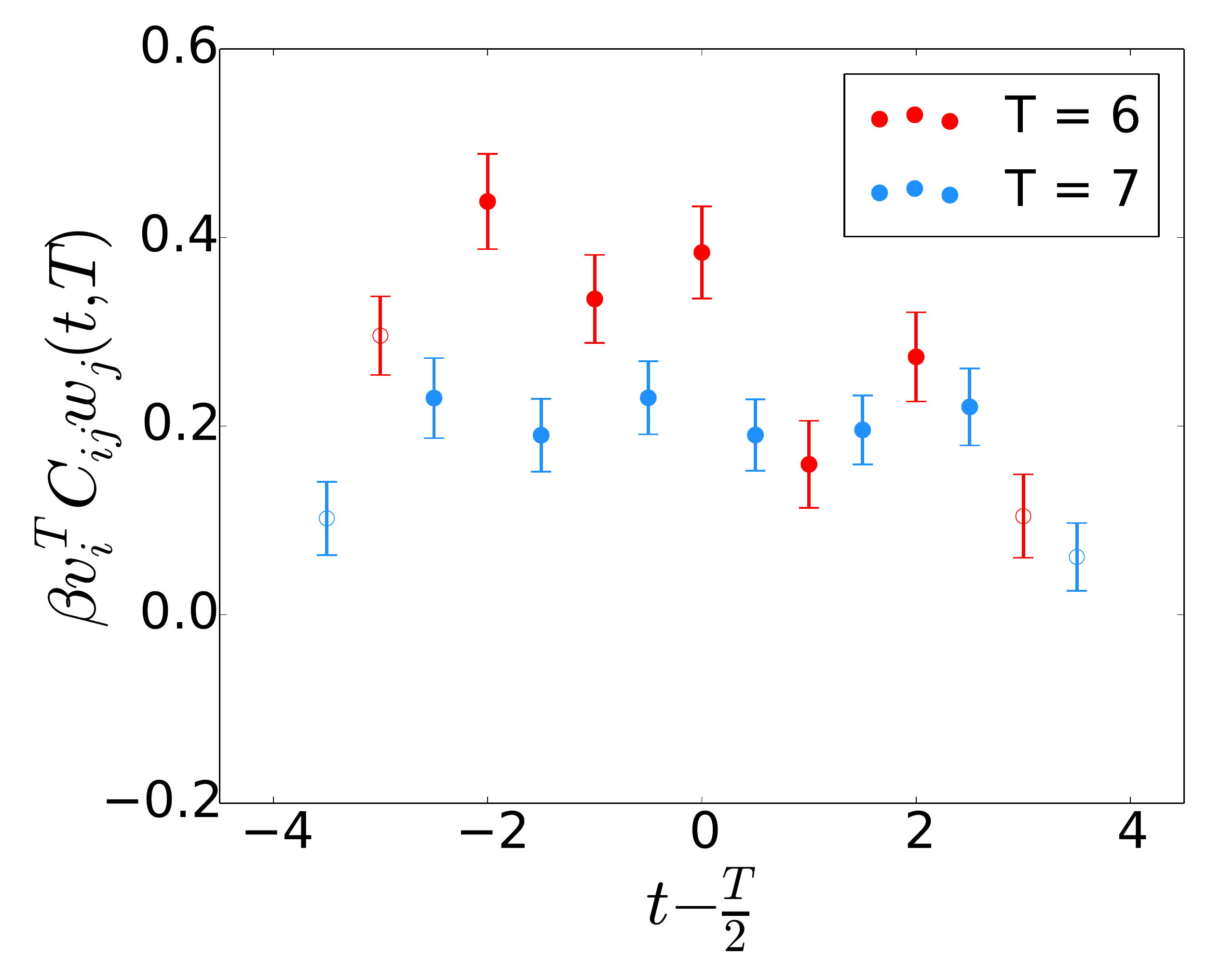}};
\node at (-1.1,2.2) {\textcolor{red}{\footnotesize PRELIMINARY}};
\end{tikzpicture} \\
\end{array}
$
\caption{
 \label{fig:3ptopt}
 Example $g_A$ three-point correlation functions for source-sink separation $T$
  and current insertion time $t$.
 The lattice ensemble used is the same as in Sec.~\protect\ref{sec:2pt},
  but with half the number of measurements.
 The red points are the data for $T=6a$ and the blue are for $T=7a$.
 Left: Raw data with no optimization.
 Right: The optimized correlation function.
}
\end{figure}

For these proceedings, we study only the raw three-point function at zero momentum.
We can again apply the signal to noise optimization with the expression
\begin{equation}
\frac{S^2}{N^2} = \sum_{ij}\sum_{\tau=1}^{t-1}
 \frac{\left[v_i C_{ij}(t,T) w_j\right]^2}
 {\delta\left[v_i C_{ij}(t,T) w_j\right]^2}\,,
\end{equation}
where the source-sink separation $T$ is held fixed and the current insertion times
 $t$ are summed over.
The three-point functions before and after the optimization are shown in Fig.~\ref{fig:3ptopt}.

\section{Outlook and Conclusions}
\label{sec:outlook}

The computation of the axial form factor using staggered quarks offers
 a new approach for addressing the $g_A$ puzzle.
Despite simplifications in our analysis,
 we have demonstrated that we can disentangle more excited states
 than what is implied by the variational method alone.
We expect our precision to improve with a more sophisticated analysis.
We have USQCD resources for computing inversions on the
 $a\approx 0.12$ and $0.09~{\rm fm}$ lattice ensembles
 and enough propagators to increase statistics by a factor of 2 have already been computed.
We plan to include a full error budget in our final analysis and will remove the blinding when
 it is done.

The axial form factor is a key component of the free nucleon
 cross section which is essential to the study of neutrino oscillations.
This form factor is a dominant contribution to systematic errors in the
 cross section, and the dipole shape ansatz severely underestimates the form factor error.
To ensure proper understanding of systematic errors on cross sections from theory,
 we plan to use the $z$~expansion parametrization.
The $z$~expansion has been successful in $B$ meson
 physics~\cite{Lattice:2015tia,Lattice:2015rga,Flynn:2015mha}
 as well as nucleon physics~\cite{Bhattacharya:2011ah,Meyer:2016oeg},
 and we plan to extend its success into the study of neutrino oscillation physics.
Some key aspects of our work -- the $z$~expansion and even physical-mass ensembles --
 are not unique to staggered quarks and the HISQ ensembles.
We can anticipate that other lattice collaborations will join us in
 aiding future neutrino experiments.

\section{Acknowledgments}
Computation for this work was done on the USQCD facilities at Fermilab and,
 for the MILC ensembles, at the Argonne Leadership Computing Facility,
 the National Center for Atmospheric Research, the National Center for Supercomputing Resources,
 the National Energy Resources Supercomputing Center, the National Institute for Computational
 Sciences, the Texas Advanced Computing Center, and under grants from the NSF and DOE.
A.S.M.\ and A.S.K.\ thank the Kavli Institute for Theoretical Physics, which is supported by the
 National Science Foundation under Grant No.~PHY11-25915, for its hospitality.
This work was supported by the U.S.\ Department of Energy SCGSR program and Universities
 Research Association (A.S.M.); the German Excellence Initiative,
 the European Union Seventh Framework Programme,
 and the European Union's Marie Curie COFUND program (A.S.K.).
The SCGSR program is administered by the Oak Ridge Institute for Science and Education for the
 DOE under contract No.\ DE-AC05-06OR23100.
R.J.H.\ and A.S.M.\ also supported by DOE Grant No.\ DE-FG02-13ER41958.
TRIUMF receives federal funding via a contribution agreement with the National
 Research Council of Canada.
Research at Perimeter Institute is supported by the Government of Canada through the
 Department of Innovation, Science and Economic Development and by the Province of Ontario
 through the Ministry of Research and Innovation.
R.L.\ thanks Intel$\textsuperscript{\textregistered}$ for its support of the Intel
 Parallel Computing Center at Indiana University.
Fermilab is operated by Fermi Research Alliance, LLC, under Contract No.\ DE-AC02-07CH11359 with
 the United States Department of Energy.

\end{document}